\documentstyle[12pt]{article}
\begin{document}
\begin{center}
\large{ \bf Coherent states for the anharmonic oscillator \\
and classical phase space trajectories}\\
\vskip 3em
{\bf H.S.Sharatchandra}\\
{\bf  The Institute of Mathematical Sciences, \\
Taramani P.O., Chennai 600-113, India}
\end{center}

Electronic address: sharat@imsc.ernet.in
\vskip 9em
\noindent{\bf Abstract}
Unique set of coherent states for the anharmonic oscillator 
is obtained by requiring i.\ under the quantum mechanical
time evolution  a coherent state evolves into another,
governed by trajectory in the classical phase space (of
a related hamiltonian); ii.\ the resolution of identity involves
exactly the classical phase space measure.The rules are
invariant under unitary transformations of the quantum 
theory and canonical transformations of the classical
theory. The states are almost, but not quite, minimal uncertainty
wave packets. The construction can be generalized to 
quantum versions of integrable classical theories.

\newpage
Coherent states for the quantum  harmonic oscillator
are defined by,
\begin{eqnarray} 
|z)&=& e^{-\overline z z/2}\sum_{0}^{\infty} \frac{z^n}
{\sqrt {n!}} |n> 
\label{1}
\end{eqnarray} 
\noindent 
They were constructed by Schr\"{o}dinger \cite{schr} as quantum 
states which have close relation to classical dynamics:
A.Expectation values $\hat q$ and $\hat p$ of position and momentum
operators in the state $|z)$ are related to the label $z$ by
\begin{eqnarray} 
z&=&\sqrt \frac{\omega}{2\hbar} q +i\sqrt \frac{1}{2\hbar \omega} p
\label{2} 
\end{eqnarray} 
\noindent 
Altenatively this may be viewed as providing an one-to-one 
correspondence between the coherent states
and the classical phase space.
B.Also a coherent state evolves into another coherent state
under time evolution with the label $z$ following the 
classical trajectory:
\begin{eqnarray} 
|z)& \rightarrow & |z(t)), \;  z(t)=ze^{-i\omega t}
\label{3} 
\end{eqnarray} 
\noindent 

Coherent states have since been rediscovered and found relevant
in different contexts \cite{klau,klau1}.There are also 
generalizations of the concept for some other quantum systems
\cite{klau,per,klau1}.These generalizations are
mostly based on certain other features of the `canonical'
coherent states defined above.
i.They are minimum uncertainty states of position and momentum.
ii.They are eigenstates of the non-Hermitian annihilation operator.
iii.They are obtained by the action of an element of the Heisenberg
group on the fiducial state $|0>$.
iv.They provide a representation of the Hilbert space by
entire functions.

The original motivation of Schr\"{o}dinger of making contact 
with classical dynamics has become very relevant through 
experimental studies \cite{ryd} of microwave ionization of hydrogen
Rydberg states and through manifestations of classical chaos
in quantum mechanics \cite{eck}.In this context there are many 
proposals for coherent states. (Ref. \cite{klau3} and references
therein.) In particular Klauder \cite{klau3} has proposed coherent 
states for the hydrogen atom by requiring its evolution into 
another coherent state.However the coefficients are largely left
unspecified or are fixed by requiring minimal uncertainty.  
Relation to classical phase space or dynamics is not imposed.

Here we propose a set of rules that provide a  unique set coherent
states for the quantum anharmonic oscillator, one  which has exact 
relation to the classical phase space and dynamics and a precise
semi-classical connection.Our rules are invariant under
unitary transformations of the quantum theory and canonical
transformations of the classical theory.  Our criteria can
be readily applied to other quantum systems.In particular,
they provide a unique set of coefficients for the hydrogen atom.  
This and its relation to the 
earlier proposals will considered elsewhere.

For our construction, in addition to the properties A and B,
the following properties of the canonical coherent 
states are crucial.

C.Semiclassical limit: The coefficients peak at the value $n$ 
where \\
$d/dn (nln|z|^2-ln n!)=0$.
This gives the largest coefficient to be the one corresponding 
to $n \approx |z|^2$.(Width of the peak is $O(|z|)$).

Consider Bohr-Sommerfeld quantization condition 
$\oint p dq=(n+1/2)h$.Now,
\begin{eqnarray} 
\int_0^T p \dot{q}dt =\frac{1}{2}a^2\omega^2 T
\label{5}
\end{eqnarray} 
\noindent 
where the amplitude $a=\omega^{-1}\sqrt{p^2+\omega^2q^2}$.  Thus 
$(p^2+\omega^2q^2)/2=(n+1/2) \hbar \omega$. With the identification
Eqn. \ref{2}, this is precisely where the coefficients peak.
Therefore the relation to the classical phase space is
more than just an identification of the labels.
For large $|z|$, the canonical coherent state is dominated by 
stationary states near the one corresponding to the Bohr 
quantization of the corresponding classical orbit. 

It is to be noted that the properties A and B would be true 
with any choice of the n-dependence of the coefficients.
The property C fixes the n-dependence atleast asymptotically.

D.Resolution of the identity: We have,
\begin{eqnarray} 
\bf 1&=&\int \frac{d^2z}{\pi} |z)(z|
\label{6}
\end{eqnarray} 
\noindent 
Now with the identification Eqn.\ref{2},
$\pi^{-1}d^2 z=h^{-1} dp dq$ is the 
canonical measure  on the classical phase space
measured in units of the Planck's constant.$z$ and $n$ 
dependence of the coefficients as in Eqn. \ref{1} is crucial 
for this simple interpretation. For instance, if we had 
\begin{eqnarray} 
|z)&=& \sum_{0}^{\infty} c_n(r)e^{-in\theta} |n>,  
\, \, \sum_{0}^{\infty} |c_n(r)|^2 =1 
\label{7}
\end{eqnarray} 
\noindent 
where $z=r exp(-i\theta)$, then properties A and B would
be true and C could also be valid for a wide choice of $c_n(r)$
(for e.g., any monotonically increasing  function 
of $r^n/\sqrt {n!}$).States $|z)$ would still provide an
(over)complete set of states and therefore a resolution of the
identity would be still possible.Now,
\begin{eqnarray} 
\int_0^{2\pi} \frac {d\theta}{2\pi}|z)(z| =\sum_{0}^{\infty} 
|c_n(r)|^2 |n><n| 
\label{8}
\end{eqnarray} 
\noindent
For a resolution of the identity we need a measure $d\mu(r)$
such that

\begin{eqnarray} 
\int d \mu(r) |c_n(r)|^2 =1
\label{9}
\end{eqnarray} 
\noindent
As has been emphasised by Klauder \cite{klau3},There are many 
choices of $c_n(r)$ meeting this requirement.
But the measure $\int d \mu(r)d \theta/\pi$ 
would not have a simple interpretation as the canonical measure 
on the classical phase space.The set of coefficients of 
the canonical coherent states are unique in this respect.

We use the ingredients A-D to construct coherent states for other
quantum systems.We first consider the Hamiltonian 
\begin{eqnarray} 
H=(a^*a+1/2)\hbar \omega +\lambda((a^*a+1/2) \hbar \omega)^2
\label{10}
\end{eqnarray} 
\noindent
which is already in the diagonal form.Relevance to the 
anharmonic oscillator will become clear later.Corresponding
classical Hamiltonian is $H((p^2+ \omega^2 q^2)/2)$
obtained by the replacement
\begin{eqnarray} 
(a^*a+1/2) \hbar \omega =(p^2+ \omega^2 q^2)/2
\label{11} 
\end{eqnarray} 
\noindent 
Classical equations of motion are
\begin{eqnarray} 
\dot{q}=H'p,\; \dot{p}=-\omega^2H'q
\label{12} 
\end{eqnarray} 
\noindent 
where $H'(y)=dH(y)/dy$ is a constant of motion related to
the amplitude or energy.The classical trajectories 
in the phase space are circles as in the harmonic case but the 
frequency $\Omega=H'\omega$ depends on the amplitude.
Classical trajectories are
\begin{eqnarray} 
q=R \; cos(\Theta+\omega H't),\, p=-R\omega \; 
sin(\Theta+\omega H't)
\label{13} 
\end{eqnarray} 
\noindent 
where the amplitude 
$R=R(t)=\omega^{-1}\sqrt{p(t)^2+ \omega^2 q(t)^2}$
is a constant of motion.  We may regard $R(t)$
as the action variable and $\Theta(t)=tan^{-1}(p(t)/(\omega q(t)))$
as the conjugate angle variable.Their 
time evolution is given by
\begin{eqnarray} 
R(t)=constant,\,  \Theta(t)= \Theta+\omega H't,
\label{14} 
\end{eqnarray} 
\noindent
In order to construct the coherent states now we now use 
$R$ and $\Theta$ to label the classical phase space: 
\begin{eqnarray} 
|R,\Theta)&=& \sum_{0}^{\infty} c_n(R,\Theta) |n>,  
\label{15} 
\end{eqnarray} 
\noindent
where the energy eigenstates $|n>$ are now the same as in the
harmonic case, but have different energy eigenvalues as the 
Hamiltonian is different.  We now apply our criteria 
A-D for the coefficients $c_n(R,\Theta)$.
Under time evolution, $|n>\rightarrow exp(-iE_nt/ \hbar)|n>$ where
$E_n$ are the exact eigenvalues of $H$ obtained by replacing 
$a^*a$ by $n$ in  Eqn. \ref{10}.
In order that a coherent state evolve into a coherent state
the only possible choice is
\begin{eqnarray} 
|R,\Theta)&=& \sum_{0}^{\infty}
 c_n(R)exp(-iE_n\frac{\Theta}{\hbar \omega H'}) |n>
\label{16} 
\end{eqnarray} 
\noindent
This brings home the crucial difference with the harmonic case.
The angle variable $\Theta$ in the classical phase space 
is periodic with range $[0, 2\pi)$.But the quantum mechanical
state $|R,\Theta)$ defined above is not periodic in $\Theta$
when the energies $E_n$ are incommensurate.This is unavoidable
once we impose criterion B as has been noted by Klauder
\cite{klau3}.Thus the label $\Theta$ has to be extended to the
covering space, $\Theta \epsilon (-\infty, \infty)$.
In spite of this the relation to the classical phase
space with $\Theta$ as the angle variable
survives, as seen below.For an observable $O$, time dependence of 
the expectation value is \\
$<O(t)>=\sum <m|O|n>exp(-i(E_n-E_m)t/\hbar)$
which can never be periodic in case the energy levels are 
incommensurate with each other and the state is not stationary.
Instead, it is an {\sl almost periodic function} \cite{boh}
of time.  There is no 
possibility of relating it to periodic orbits except in the sense 
considered above and therefore this feature need not be regarded
as undesirable.

We now apply criterion C.  The Bohr quantization gives
\begin{eqnarray} 
\frac{1}{2}R^2\omega^2 H'T=nh
\label{17}
\end{eqnarray} 
\noindent 
Now the period $T=2\pi (\omega H')^{-1}$.Therefore we get the 
quantization condition $(p^2+\omega^2q^2)/2=(n+1/2)\hbar \omega$ 
exactly as in the harmonic case
with $a^*a \rightarrow n$ as is to be expected.

For criterion C we require the coefficients $c_n(R)$ to peak 
at $n \approx R^2$.Same choice as in the harmonic oscillator
suffices:$c_n(R)=R^n/\sqrt{n!}$.However, 
the choice is not unique at this stage.

We now apply criterion D.As $|R,\Theta)$ is not periodic in 
$\Theta$ we cannot simply integrate over the range $[0,2\pi)$
in order to resolve the identity.In order to retain the 
interpretation of the canonical phase space measure, we will
consider averaging over infinitely many classical orbits:
\begin{eqnarray} 
[\int_{-\pi}^{\pi}] \frac{d \Theta}{2\pi}=lim_N \rightarrow \infty 
\; \;\frac{1}{N}\int_{-\pi N}^{\pi N} \frac{d \Theta}{2\pi}
\label{19}
\end{eqnarray} 
\noindent 
We now have
\begin{eqnarray} 
[\int_{-\pi}^{\pi}] \frac{d\Theta}{2\pi} 
e^{-i(E_n-E_m)\Theta}=\delta_{nm}
\label{20}
\end{eqnarray} 
\noindent 
Therefore
\begin{eqnarray} 
[\int_{-\pi}^{\pi}]|R,\Theta)(R,\Theta|=\sum_0^{\infty}|c_n(R)|^2 
|n><n| 
\label{21}
\end{eqnarray} 
\noindent
This shows that the choice also satisfies the criterion D.
\begin{eqnarray} 
\bf 1&=&\int_0^{\infty} \frac{dR^2}{2} [\int] \frac{d\Theta}{\pi} 
|R,\Theta)(R,\Theta|
\label{22}
\end{eqnarray} 
\noindent 
with $(2\pi)^{-1}dR^2 d\Theta =h^{-1}dp dq$. 
Thus we have got a unique set of coherent states closely 
related to the classical phase space and time evolution.

The form of Eqn. \ref{16} suggests a new choice of conjugate 
variables with $\tau=(\omega H')^{-1}\Theta=
(\omega H')^{-1}tan^{-1}(p/(\omega q))$
as the new angle variable.The conjugate action variable
is just the Hamiltonian $H$ as in the Hamilton-Jacobi theory.
Under time evolution we have,
\begin{eqnarray} 
H=constant, \; \tau \rightarrow \tau+t
\label{23}
\end{eqnarray} 
\noindent 
$\tau$ is the time variable promoted to a dynamical degree
of freedom.As this is a canonical set, the phase space measure
is simply $ dH d\tau$.The range of $\tau$ is
$[0,2\pi/(\omega H'))$ (one period), and not just $[0,2\pi)$, and
therefore depends on the value of $H$.We could label the
coherent states $|R,\Theta)$ by $|H,\tau)$ as well,
with the understanding that $R$ in $c_n(R)$ is reexpressed 
in terms of $H$. The coefficients ofcourse would have different
forms for different Hamiltonians, and not have the simple form
as when written in terms of the $R$ variable.  On the other hand, 
the phase has a simple form $exp(-iE_n \tau/ \hbar)$.

We now consider the anharmonic oscillator 
\begin{eqnarray} 
{\hat {\bf H}}=\frac{1}{2}(\hat{P}^2+\omega^2 \hat{Q}^2 
+\lambda \hat{Q}^4)
\label{24}
\end{eqnarray} 
\noindent
When $\lambda \neq 0$ the orbits in the phase space are closed 
but not circles and has an amplitude dependent frequency.
By a canonical transformation we may pass to the action angle 
variables.Now the orbits are circular.In the quantum mechanical
case this procedure corresponds to a unitatry transformation
that diagonalises the Hamiltonian:
\begin{eqnarray} 
{\hat{\bf H}}(\hat{P},\hat{Q})=
 \hat{H}((\hat{p}^2+\hat{q}^2)/2);\\ 
\hat{P}(\hat{p},\hat{q})=\hat{U}\hat{p}\hat{U}^*, \; 
\; \hat{Q}(\hat{p},\hat{q})=\hat{U}\hat{q}\hat{U}^* 
\label{25}
\end{eqnarray} 
\noindent
$H(y)$ is a monotonic function of the argument 
such that $H(y=(n+1/2) \hbar)=E_n$, the energy levels.
In terms of the creation and annihilation operators $a^*,a$
for $p,q$, a formal expression for the diagonalised form  is 
$H=\sum_0^{\infty}H_n a^{*n}a^n$, where the coefficients $H_n$
are related to the energy levels $E_n$ {\sl via} 
\begin{eqnarray} 
H_n=\frac{1}{n!}\frac{d^n}{dy^n}(e^{-y}\sum_0^\infty
\frac{E_m}{m!}y^m )|_{y=0}
\label{29}
\end{eqnarray} 
\noindent

For the Hamiltonian $H$ we may simply apply our construction of 
the coherent states.These will also be the coherent states of the
anharmonic oscillator.$|n>$ are now the energy eigenstates of the
anharmonic oscillator, labelled by  non-negetive integers.
The coherent states are labeled by the conjugate pair
$R$ and $\Theta$ (or equivalently, $H$ and $\tau$) defined in
terms of the pair $p,q$ as before. We would like to
label it by the conjugate pair $P$ and $Q$ in the Hamiltonian
Eqn. \ref{24}.But there would be no canonical transformation of 
the classical theory which takes the Hamiltonian
$H((p^2+q^2)/2)$ to the Hamiltonian ${\bf H}(P,Q)$, Eqn.\ref{24}.
Only in the region $(p^2+q^2)>> \hbar$ would the two be related
by a canonical transformation. We could in any case choose the 
new conjugate pair ${\cal P},{\cal Q}$ to get the Hamiltonian in 
the form ${\cal H}={\cal P}^2/2+u({\cal Q})$ with a new 
potential $u({\cal Q})$.In the present case it is easy to give the
formal method for obtaining the new potential $u({\cal Q})$ and
the relationship between the pairs $H,\tau$ and 
${\cal P},{\cal Q}$. We need to only match energies and periods of
the orbits of the two systems.Explicitly,
\begin{eqnarray} 
\frac{\pi}{H'}= 2\int_0^{\cal H}du\frac{d{\cal Q}/du}
{\sqrt{2(H-u)}}
\label{26}
\end{eqnarray} 
\noindent
matches the periods of an orbit in terms of the two sets of 
variables.This may be viewed as an implicit functional equation for
$\cal{Q}$ as a function of $u$ or equivalently, the new potential 
$u(\cal{Q})$ in terms of $H$.Once  $u(\cal{Q})$ is known, 
$\cal{P}$ can be obtained from
\begin{eqnarray} 
H=\frac{1}{2}{\cal P}^2+u({\cal Q})
\label{27}
\end{eqnarray} 
\noindent
which is a matching of the energies in the two variables.
Finally $\cal{Q}$ is related to the old pair $H,\tau$ by
\begin{eqnarray} 
\tau=\int_0^{\cal Q} \frac {d{\cal Q}}{\sqrt{2(H-u(\cal{Q}))}}
\label{28}
\end{eqnarray} 
\noindent
which uses the interpretation of $\tau$ as the canonical variable
corresponding to time. When labelled in terms of $\cal{P}$ and 
$\cal{Q}$, the time evolution of the coherent states will be 
given by the classical trajectories of  the Hamiltonian with the 
potential $u(\cal{Q})$ and not of the anharmonic oscillator we 
started with. $\tau$ would be a multivalued function of $\cal{P}$ 
and $\cal{Q}$, and it is important that this multivaluedness be 
retained in the coherent state written in terms of $\cal{P}$ and 
$\cal{Q}$.

Canonical coherent states are minimal uncertainty wave packets
with the expectation value of position and momentum directly
given by the imaginary and the real part respectively of the label
$z$. We now consider these issues for our coherent states.
We first discuss the case of the
Hamiltonian Eqn. \ref{10}.When the energy levels are 
incommensurate relative to each other, these coherent states can
no longer be eigenfunctions of the annihilation operator $a$.
We have,
\begin{eqnarray} 
a|R,\Theta)&=& e^{-R^2}\sum_{0}^{\infty}
\frac {R^{n+1}}{\sqrt{n!}}exp(-i(E_{n+1}-E_n)\frac{\Theta}
{\hbar H'}) 
\label{30} 
\end{eqnarray} 
\noindent
Only for the harmonic oscillator, we get $R \; exp(-i\Theta)$ on 
the r.h.s.In the general case this is an almost periodic function
of time \cite{boh}.In particular the value gets arbitrarily 
close  but does not quite repeat after an interval of time.However,
for large $R$, the coefficients $n \approx R^2$ dominate
and $(E_{n+1}-E_n)\approx \hbar H'$ for such values.  Therefore 
$a|R,\Theta) \approx R exp(-i \Theta)|R,\Theta)$(By this we mean
$(a-R exp(-i \Theta))|R,\Theta)$ has a small norm.)
In particular this means that expectation values of $p$ and $q$
in our coherent states $|R,\Theta)$ are approximately 
$R \; sin \Theta$ and $R \; cos \Theta$, the approximation becoming
rapidly better for large $R$.

We now consider the spread in position and momentum and Heisenberg 
uncertainty for our coherent states. An explicit calculation gives,
$\Delta q^2=\hbar (2\omega)^{-1}(1+R^2 k(R,\Theta))$, where
\begin{eqnarray*} 
k(R,\Theta)&=& 2e^{-R^2}\sum_{0}^{\infty}
\frac {R^{2n}}{n!}(cos \frac{(E_{n+2}-E_n)\Theta}{\hbar
H'\omega}-cos (2\Theta)) \\
& &- 4e^{-R^2}\sum_{0}^{\infty} \frac {R^{2n}}{n!}
( cos \frac{(E_{n+1}-E_n)\Theta}{\hbar H'\omega}- cos \Theta) \\
& &\times (2+e^{-R^2}\sum_{0}^{\infty}
\frac {R^{2n}}{n!}(cos \frac{(E_{n+1}-E_n)\Theta}{\hbar
H'\omega}- cos \Theta)
\label{40} 
\end{eqnarray*} 
\noindent
$k(R,\Theta) \rightarrow 0$ for $R \rightarrow \infty$.If this
asymptotic falloff is faster than $R^{-2}$, then 
$\Delta q^2 \approx \hbar(2\omega)^{-1}$. In the present 
case an explicit asymptotic analysis can be made to justify this
result.Similar results are also valid for $\Delta p^2$.
In particular this means that atleast for large
$p^2+q^2$, the coherent state is almost a minimal uncertainty 
wave packet, the correction being very tiny for large $R$.

We now argue that these properties are also valid  for the 
observables ${\hat P},{\hat Q}$ of the anharmonic oscillator,
Eqn.\ref{24}.${\hat p},{\hat q}$ are operator expressions in 
${\hat P},{\hat Q}$ such that when substituted in 
 ${\hat H}(({\hat p}^2+{\hat q}^2)/2)$ only dependence on
$P^2$ is quadratic, Eqn.\ref{25}. On the other hand, the
classical variables $p,q$ are functions of
${\cal P},{\cal Q}$ such that when substituted in the 
classical Hamiltonian $H((p^2+q^2)/2)$, we get  
${\cal P}^2/2+u({\cal Q})$, with only a quadratic term in
${\cal P}$.The two sets of functions differ only in 
higher orders in $\hbar$, in particular, due to 
ordering of operators.In the semiclassical region,
of large $p^2+q^2$, these are neglegible.  In this region
the expectation values $(R,\Theta|{\hat P}|R,\Theta)$ and
$(R,\Theta|{\hat Q}|R,\Theta)$ are computed by substituting
the operator expressions in ${\hat p},{\hat q}$, rewriting 
these expressions in normal ordered form in $a,a^*$ and
simply replacing $a \rightarrow R \; exp(-i\Theta)$ and
$a^* \rightarrow R \; exp(+i\Theta)$. Therefore upto 
terms of $O(h)$, these are same as $\cal P$ and $\cal Q$
for these values of $R$ and $\Theta$.Also as the 
canonical transformations preserve the area, the spread
of the wavepacket in $\cal{P},\cal{Q})$ is the same as the 
spread in $p,q$.  Thus they are {\sl almost} minimum 
uncertainty states, though not exactly so.
Therefore our coherent states for the 
anharmonic oscillator have properties of the canonical 
coherent states in the semiclassical region.

The coherent states we have constructed are same for 
Hamiltonians that are equivalent under unitary transformations.
They may be labelled by any choice of conjugate variables of the
classical phase space.Their properties $\sl vis-a-vis$ classical
dynamics are invariant under the choice of the conjugate
variables.

Our construction uses the one parameter group associated with 
the time evolution crucially and is not tied to other groups such
as the Heisenberg group or symmetry groups of specific 
Hamiltonians, such as the rigid rotator or the hydrogen atom.As
such it can be used for general Hamiltonians.This will be 
considered elsewhere cite \cite{ms}.

I thank Professor K.Mariwalla for critical comments and 
Professor R.Simon for clarifying the issues involved and
for correcting some errors.I also thank Pushan
Mazumdar for a careful reading of the manuscript.


\begin{thebibliography}{99}

\bibitem{schr}
E. Schr\"{o}dinger,  {\sl Naturwissenschaften } 
{\bf 14}, 644(1926).
\bibitem{klau}
J.R.Klauder  and B-S. Skagerstam, {\sl Coherent States:
Applications in Physics and Mathematical Physics} 
(World Scientific, Singapore, 1985).
\bibitem{klau1}
{\sl Coherent States: Past, Present and Future } 
eds. D.H.Feng, J.R.Klauder  and M.R.Strayer, 
(World Scientific, Singapore, 1994).
\bibitem{per} A.M.Perelomov,
{\sl Generalized Coherent States and Their Applications} 
(Spinger Verlag, Berlin, 1986).
\bibitem{ryd}R.F.Stebbings and F.B.Dunning,
{\sl Rydberg States of Atoms and Molecules}
(Cambridge Univ. Press, London,1983).
\bibitem{eck}
B.Eckhardt, {\sl  Phys. Rep.} 
{\bf 163},205(1988).
\bibitem{klau3}
J.R.Klauder, {\sl J. Phys. A: Math. Gen.} 
{\bf A29},L293-298 (1996).
\bibitem{boh}A.S.Besikovitch,
{\sl Almost Periodic Functions}
(Dover Publications Inc.,1954).
\bibitem{ms}
Pushan Majumdar and H.S.Sharatchandra, under preparation.

\end{thebibliography}
\end{document}